\documentclass[aps,prb,preprint,superscriptaddress,showpacs,floatfix]{revtex4}
\usepackage{graphicx,epsfig,amsmath}

\newcommand{\kboltz}{\mbox{$k_{B}$}}

\newcommand{\Vds}{\mbox{$\text{V}_{\text{ds}}$}}
\newcommand{\Vg}{\mbox{$\text{V}_{\text{g}}$}}

\begin{document}


\title{Charge Transport in Arrays of PbSe Nanocrystals}

\author{T. S. Mentzel}	
	\email{tamarm@mit.edu}
	\affiliation{Department of Physics, Massachusetts Institute of Technology, Cambridge, Massachusetts 02139}
\author{V. J. Porter}
	\affiliation{Department of Chemistry, Massachusetts Institute of Technology, Cambridge, Massachusetts 02139}
\author{S. Geyer}
	\affiliation{Department of Chemistry, Massachusetts Institute of Technology, Cambridge, Massachusetts 02139}
\author{K. MacLean} 
	\affiliation{Department of Physics, Massachusetts Institute of Technology, Cambridge, Massachusetts 02139}
\author{M. G. Bawendi}
	\affiliation{Department of Chemistry, Massachusetts Institute of Technology, Cambridge, Massachusetts 02139}
\author{M. A. Kastner} 
	\affiliation{Department of Physics, Massachusetts Institute of Technology, Cambridge, Massachusetts 02139}


\begin{abstract}
We report electrical transport measurements of arrays of PbSe nanocrystals forming the channels of field effect transistors.  We measure the current in these devices as a function of source-drain voltage, gate voltage and temperature. Annealing is necessary to observe measurable current after which a simple model of hopping between intrinsic localized states describes the transport properties of the nanocrystal solid.  We find that the majority carriers are holes, which are thermally released from acceptor states.  At low source-drain voltages, the activation energy for the conductivity is given by the energy required to generate holes plus the activation over barriers resulting from site disorder. At high source-drain voltages the activation energy is given by the former only. The thermal activation energy of the zero-bias conductance indicates that the Fermi energy is close to the highest-occupied valence level, the 1S$_{h}$ state, and this is confirmed by field-effect measurements, which give a density of states of approximately eight per nanocrystal as expected from the degeneracy of the 1S$_{h}$ state. 

\end{abstract}


\maketitle


\section{Introduction}

Colloidal semiconductor nanocrystals can be made to self assemble into a close-packed array, creating a novel material known as a nanocrystal (NC) solid.\cite{Bawendi:packing, Heath:packing, Markovitch, Murray:SynthesisII, Redl, Shevchenko}  Electrons in an NC solid made from semiconductor NCs, in contrast to metallic ones, have long-range Coulomb interactions;\cite{Orlando} therefore, the motion of electrons is expected to be highly correlated as long as the number of electrons per NC is not too small.  The ability to tune both the energy levels of the individual NCs and the electronic coupling between NCs makes these solids a promising test bed for investigating many-body physics. It has been demonstrated that the charge density in semiconductor NC solids can be modulated, \cite{Talapin:Science} which makes the system suitable for observing the predicted characteristics of electronic correlations as a function of charge density.\cite{Levitov:theory}

We study PbSe NC solids because they have higher conductances than the well-studied solids composed of II-VI NCs.\cite{Morgan:CdSe, Drndic:Anneal, Ginger, Fischbein, DrndicII, Leatherdale}  PbSe NCs display a narrower dispersion in electronic energy levels than the II-VI NCs as well as a higher degeneracy,\cite{AllanDelerue} which increases the density of states available for conduction.\cite{Optical,Talapin:Science} It is possible that charge carriers in PbSe NC solids, which are holes, are generated by thermal excitation of electrons into midgap states that arise from dangling bonds on the surface.  Bulk PbSe has midgap states closer to the band edge than bulk CdSe, and thus we expect a higher density of charge carriers in PbSe NCs.  Furthermore, PbSe NCs have attracted much attention because of their interband transitions in the IR and multiple exciton generation,\cite{Klimov} and hence potential for application in novel optoelectronic devices.\cite{LED, AlivisatosSolar, Sargent}  Such applications involve electronic transport through NC solids, thus necessitating a fundamental understanding of the conduction properties.

Experimental studies of charge transport in PbSe NC solids have revealed varied behavior.  It has been reported that this NC solid can be tuned from a regime displaying Coulomb blockade to a regime exhibiting variable range hopping with a Coulomb gap in the density of states at the Fermi energy.\cite{Romero:PRL} D. Talapin \textit{et al.} treat the NC solids with hydrazine and find variable range hopping with a constant density of states at the Fermi energy.\cite{Talapin:Science}  B. L. Wehrenberg \textit{et al.} cross-link the NCs, inject electrons into the NC films by electrochemical gating and find variable range hopping with a Coulomb gap.\cite{Wehrenberg, Wehrenberg:ChargeInjection}

We report on charge transport in an untreated PbSe NC solid that serves as the channel of an inverted field effect transistor (FET).  By modulating the voltage on the gate, we vary the charge density and find that holes are the majority carriers.   The dependence of the conduction on temperature, field and gate voltage agrees with a model in which holes are generated in the highest-occupied valence level by thermal excitation from the Fermi energy, and the Fermi energy is close to the band edge.  Our results are consistent with a model in which the intrinsic valence states are localized by site disorder, and the transport is limited by hopping between nearest neighbor localized states after thermal release of holes from acceptors. Unlike previous studies,\cite{Talapin:Science, Romero:PRL, Wehrenberg} we do not find evidence of variable range hopping or Coulomb blockade.  In some cases, the NC solids studied previously\cite{Talapin:Science, Wehrenberg} are different from ours in that they were chemically treated and electrons are the majority carriers.  Nonetheless, we find differing results even from prior studies of untreated PbSe NC solids.\cite{Romero:PRL} Section \ref{sec:exp} gives the details of our experimental methods.  Section \ref{sec:results} provides the results of our measurements, and Sections \ref{sec:disc} and \ref{sec:concl} provide discussion and conclusions respectively.

\section{Experimental Details}\label{sec:exp}

PbSe NCs are synthesized according to a modified version of a previously described method.\cite{Murray:Synthesis, Murray:SynthesisII, Steckel}	 We add 0.38 g of lead acetate trihydrate and 0.65 mL of oleic acid to 20 mL of octadecene.  The mixture is heated to 150 $^{\circ}$C under argon to dissolve the acetic acid. The solution is degassed for 2 hours at 100 $^{\circ}$C to remove acetic acid and water, thereby forming lead oleate.  The solution is heated to 180 $^{\circ}$C under argon. 5mL of 1 M solution of tri-\textit{n}-octylphosphine selenide in tri-\textit{n}-octylphosphine is injected into the solution. Lead selenide NCs nucleate upon injection. The NCs are grown for approximately 5 min and then cooled to room temperature. The resulting NCs have a diameter of 6.2$\pm$0.4 nm, as determined by TEM, with a capping layer of oleic acid of thickness $\approx$2 nm.   The solution is transferred to the nitrogen environment of a glovebox.  
	
\begin{figure} 
\setlength{\unitlength}{1cm}
\begin{center}
\includegraphics[width=8.4cm, keepaspectratio=true]{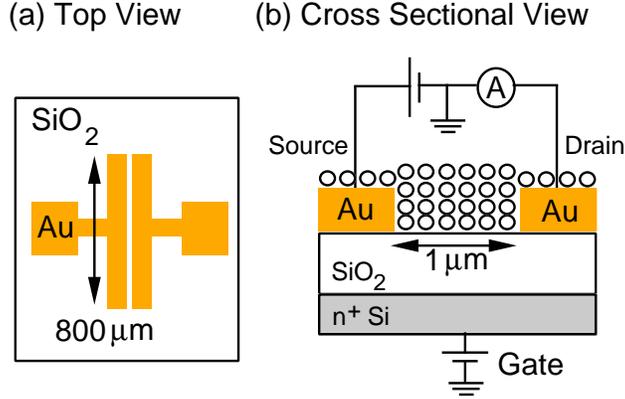}
\end{center}

\caption{Schematic of the inverted FET and the circuit used in the experiments. Silicon oxide 330 nm thick is grown on an n-doped silicon wafer. Gold electrodes 200 x 800 x 0.1 $\mu$m$^{3}$, separated by a distance of 1 $\mu$m, are microfabricated on the silicon oxide surface.  NCs 6.2 nm in diameter are drop cast from solution on the strucure, resulting in a film approximately 300 nm thick.}
\label{fig:Device}
\end{figure}

To make films of NC arrays, we process the growth solution in a glovebox according to previously reported methods. \cite{Porter:CdTe} A mixture of methanol and butanol is added to the solution to precipitate the NCs. The sample is centrifuged to collect the NCs and the supernatant is discarded.  The NCs are redissolved in hexane and passed throught a 0.2 $\mu$m filter.  The resulting solution is centrifuged to precipitate remaining salts.  The precipitate is discarded and the NCs remain in the supernatant.  The NCs are precipitated a second time in a mixture of methanol and butanol as described above except this time they are passed through a 0.1 $\mu$m filter. After precipitating the NCs a third time, the NCs are redissolved in a 9$:$1 hexane$:$octane mixture and passed through a 0.02 $\mu$m filter. The solution is drop cast onto an inverted FET as depicted in Figure \ref{fig:Device}.
   
The inverted FET [Fig. \ref{fig:Device}] consists of gold bar electrodes patterned on a silicon substrate with a 330 nm layer of thermally grown silicon oxide.\cite{Morgan:CdSe} The silicon substrate is degenerately doped with arsenic and serves as a back gate.  Pairs of gold electrodes [Fig. \ref{fig:Device}(a)] are 800 $\mu$m in length, 200 $\mu$m in width and separated by a 1 $\mu$m gap; the electrodes are 100 nm thick. The FETs are attached to a chip carrier with silver epoxy.  Gold wire bonds serve as the electrical connection between the electrodes and the chip carrier. Before depositing the films of NCs, we test the device to ensure that the leakage current between the drain and source electrodes is $<$300 fA and that the leakage current between the gate and the drain or source electrodes is $<$5 pA. We bake the bare device for 1 h at 150 $^{\circ}$C to dry the surface and deposit the film over the entire surface.  The film between the electrodes has a thickness of approximately 300 nm, corresponding to about 40 monolayers. Each monolayer has $\approx$130 NCs in series and $\approx$100 000 NCs in parallel between the electrodes.  After depositing the NCs, the solid is allowed to dry overnight in an inert atmosphere. 

The sample is transferred from a nitrogen environment to a Janis VPF-700 cryostat without exposure to air; and the cryostat is maintained under vacuum.  The sample is annealed in the cryostat at 400 - 410 K during which we monitor the increase in conductance with annealing.  After $\approx$ 35 min, the conductance saturates and we transfer the sample to an Oxford Variox cryostat. During the transfer, the sample is exposed to air for less than 5 min.  Electrical measurements are performed in the Oxford Variox cryostat while the sample is held in helium exchange gas.  For DC measurments, a Yokogawa 7611 voltage source provides the drain-source voltage, \Vds, and a Keithly 2400 sourcemeter provides voltage to the gate, \Vg. The current is amplified by an Ithaco 1211 current amplifier and is measured by an HP 34401A DMM. For measurements of differential conductance, dI/d\Vds, an HP 3325A function generator provides an AC voltage and a PARC 5316 lock-in amplifier is added at the output of the current amplifier.  The temperature is controlled with an Oxford ITC 503.

For glancing incidence small angle x-ray scattering (GISAXS) experiments, we prepare samples by baking a bare silicon substrate for 1 h at 150 $^{\circ}$C to dry the surface.  We drop cast the NCs on the silicon substrate in a nitrogen environment and the sample is not exposed to air until it is dry.  To compare the films before and after annealing, we prepare a second sample in parallel which we anneal under vacuum at 400 K for 35 min.  The oven is situated in the nitrogen environment of the glovebox so that the sample does not see air prior to annealing.  GISAXS experiments are performed with a PANalytical multipurpose diffractometer. We have measured transport properties of approximately 15 samples; for all but one the functional dependences on \Vds\space and \Vg\space are very similar.
 
\section{Results} \label{sec:results}

\begin{figure} 
\setlength{\unitlength}{1cm}
\begin{center}
\includegraphics[width=8.4cm, keepaspectratio=true]{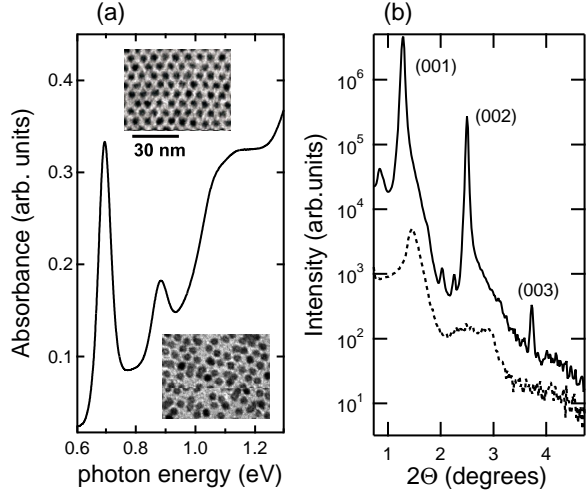}
\end{center}

\caption{(a) Optical absorption spectrum of NCs in solution.  The first absorption peak occurs at 695 meV.  The insets are TEM images of a NC film on a carbon grid before annealing (\textit{top inset}) and after annealing (\textit{bottom inset}).  The NCs have a mean diameter of 6.2$\pm$0.4 nm.  (b) GISAXS diffraction pattern of an NC film as deposited on a silicon substrate (\textit{solid line}), and after annealing at 400 K for 35 min (\textit{dotted line}). The thickness of the capping layer decreases from $\approx$2 nm to $\approx$1 nm upon annealing.}
\label{fig:AbsSAXS}
\end{figure}

Figure \ref{fig:AbsSAXS}(a) shows the optical absorption spectrum of colloidal NCs in a solution of trichloro-trifluoroethane.  The first absorption peak, corresponding to a transition that creates the lowest-energy exciton with both the electron and hole in 1S states (1S$_{h}$-1S$_{e}$),\cite{Kang, Du, Optical, Steckel} occurs at 695 meV with a full width at half maximum of 53 meV.  TEM images indicate that the NCs have an average diameter of 6.2$\pm$0.4 nm and self assemble into hexagonally close packed arrays.\cite{Steckel} Data from GISAXS [Fig. \ref{fig:AbsSAXS}(b)] show that the average thickness of the oleic acid capping layer decreases with annealing from $\approx$2 nm to $\approx$1 nm, most likely resulting from further interdigitation of the oleic acid molecule.\cite{Steckel} Upon annealing, the widths of the first two peaks broaden and the third can no longer be resolved, indicating an increase in disorder.  TEM data confirm that disorder increases with annealing.

\begin{figure} 
\setlength{\unitlength}{1cm}
\begin{center}
\includegraphics[width=8.4cm, keepaspectratio=true]{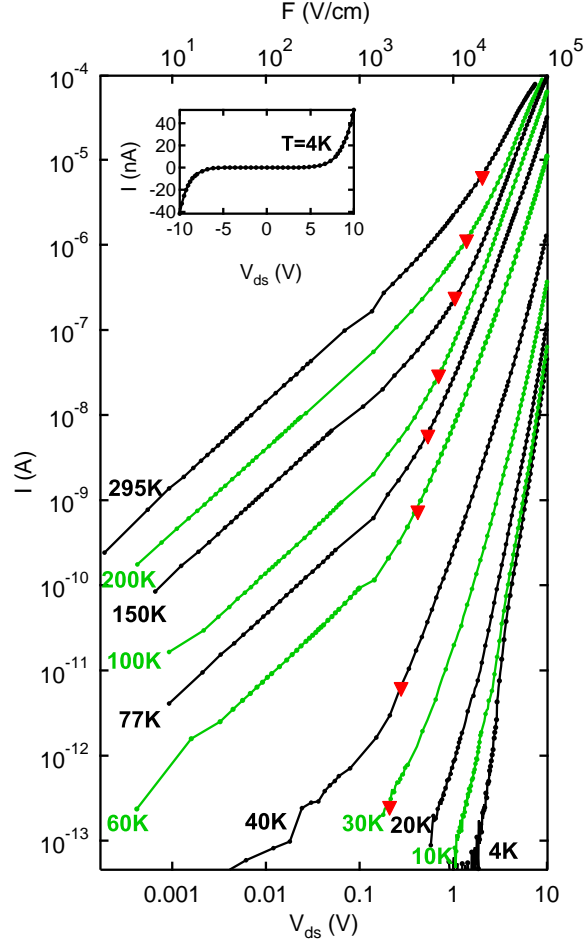}
\end{center}

\caption{Current versus drain-source voltage with \Vg\space= 0 V for various values of temperature. The lines serve as guides to the eye. The triangle marker indicates the value of \Vds\space at which a crossover from ohmic to non-ohmic behavior is predicted as explained in the text.  The top axis gives the corresponding field.  The inset is a plot of current versus \Vds\space at 4 K and \Vg = 0 V.}
\label{fig:TempFull}
\end{figure}

Prior to annealing, the current through the array is immeasurable ($<$ 100 fA) with an applied field up to 10$^{5}$ V/cm.
After the film is annealed at 400 K for 35 min, the zero-bias conductance at 295 K is 10$^{-6}$ $\Omega^{-1}$, an increase of more than six orders of magnitude.  Figure \ref{fig:TempFull} displays the current in the annealed film as a function of drain-source voltage with \Vg\space= 0 for different temperatures. We observe ohmic behavior at low fields for T $\geq$ 40 K. In this regime, the temperature dependence is strongest.  For T $<$ 40 K as well as in a high-field regime for T $\geq$ 40 K, the current deviates from ohmic behavior and the temperature dependence is weaker.  

Current as a function of drain-source voltage at 4 K with \Vg\space= 0 V is shown on a linear plot in the inset of Figure \ref{fig:TempFull}.  Previous studies of charge transport in arrays of PbSe NCs have found a threshold in voltage, interpreted as resulting from Coulomb blockade.\cite{Romero:PRL}  While it appears that the I-\Vds\space curve, when displayed on a linear plot in the inset of Fig. \ref{fig:TempFull}, exhibits this behavior, there is no evidence of a threshold when I-\Vds\space is plotted on a log-log scale.

\begin{figure} 
\setlength{\unitlength}{1cm}
\begin{center}
\includegraphics[width=8.4cm, keepaspectratio=true]{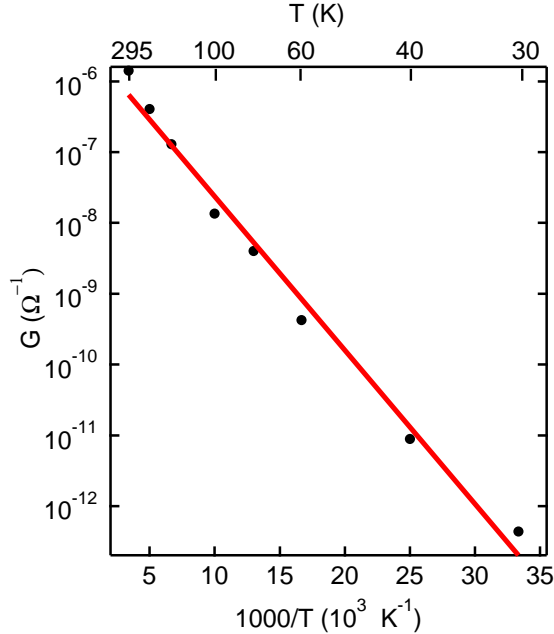}
\end{center}

\caption{Temperature dependence of the zero-bias conductance in the range 30 $<$ T $<$ 295 K with \Vg\space= 0. The solid line is an Arrhenius fit to the data with an activation energy of 42 meV.}
\label{fig:TempDep}
\end{figure}

From the data shown in Fig. \ref{fig:TempFull}, we extract the zero-bias conductance as a function of temperature and plot it in Fig. \ref{fig:TempDep}.  The zero-bias conductance is immeasurable ($<10^{-13}$ $\Omega^{-1}$) at temperatures below 30 K.  Above 30 K, the conductance fits well to an Arrehnius equation, G(T) = G$_{0}$exp[{-E$_{A}$/\kboltz T}] with an activation energy E$_{A}$ = 42$\pm$2 meV. 

\begin{figure} 
\setlength{\unitlength}{1cm}
\begin{center}
\includegraphics[width=8.4cm, keepaspectratio=true]{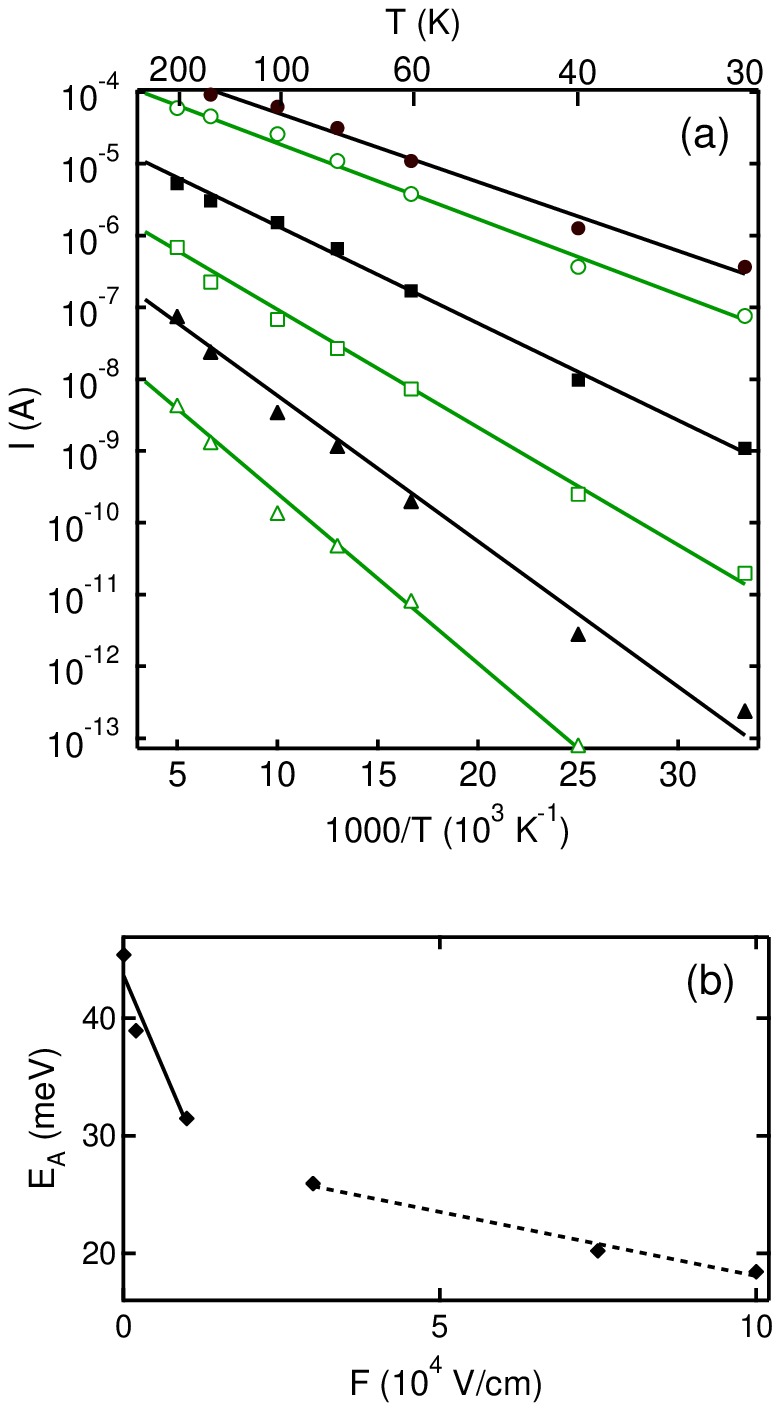}
\end{center}

\caption{(a) Temperature dependence of the current at different values of \Vds\space with \Vg=0. (filled circle \Vds\space= 10 V, open circle \Vds\space= 7.5 V, filled square \Vds\space= 3 V, open square \Vds\space= 1 V, filled triangle \Vds\space= 200 mV, open triangle \Vds\space= 10 mV). The solid lines are fits to Arrhenius equations in the range 30 K $<$ T $<$ 200 K.  (b) Activation energy, E$_{A}$, versus the field, F, established by the drain-source voltage.  The solid line is a linear fit of E$_{A}$ versus F for 10$^{2}$ V/cm $<$ F $<$ 10$^{4}$ V/cm.  The dotted line is a linear fit for 3$\times$10$^{4}$ V/cm $<$ F $<$ 10$^{5}$ V/cm}
\label{fig:VdsDependence}
\end{figure}

To explore further the temperature dependence, we plot current versus inverse temperature for various values of drain-source voltage in Fig. \ref{fig:VdsDependence}. Between 30 and 200 K, the data at each value of \Vds\space are well-described by an Arrhenius equation with an activation energy that decreases with increasing drain-source voltage.  Prior measurements of conductivity in arrays of CdTe \cite{Porter:CdTe} and PbS \cite{Scott} NCs also have found Arrhenius behavior with an activation energy that decreases with field. In Fig. \ref{fig:VdsDependence}(b), we plot the activation energy versus the applied field.

\begin{figure} 
\setlength{\unitlength}{1cm}
\begin{center}
\includegraphics[width=8.4cm, keepaspectratio=true]{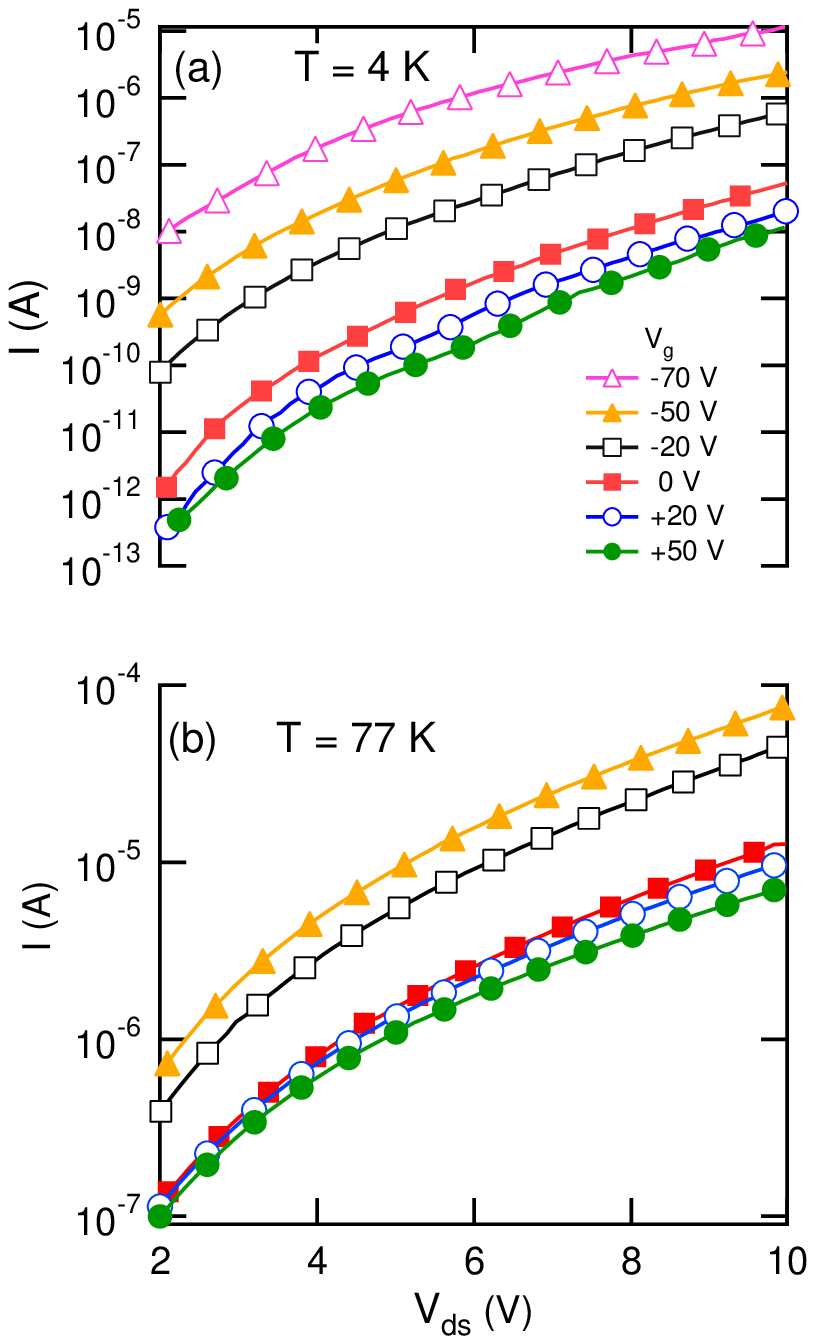}
\end{center}

\caption{Current versus \Vds\space for various values of \Vg. (a) T = 4 K  and (b) T = 77 K.  The lines serve as guides to the eye.}
\label{fig:VgDC3}
\end{figure}

Figure \ref{fig:VgDC3} shows the I-\Vds\space curves at various gate voltages. When we vary \Vg\space and modulate the charge density, we find that the magnitude of the current changes, but the dependence on source-drain bias does not.  From this, we conclude that the conduction mechanism is insensitive to the charge density.  The current increases (decreases) as \Vg\space is made more negative (positive).

\begin{figure} 
\setlength{\unitlength}{1cm}
\begin{center}
\includegraphics[width=8.4cm, keepaspectratio=true]{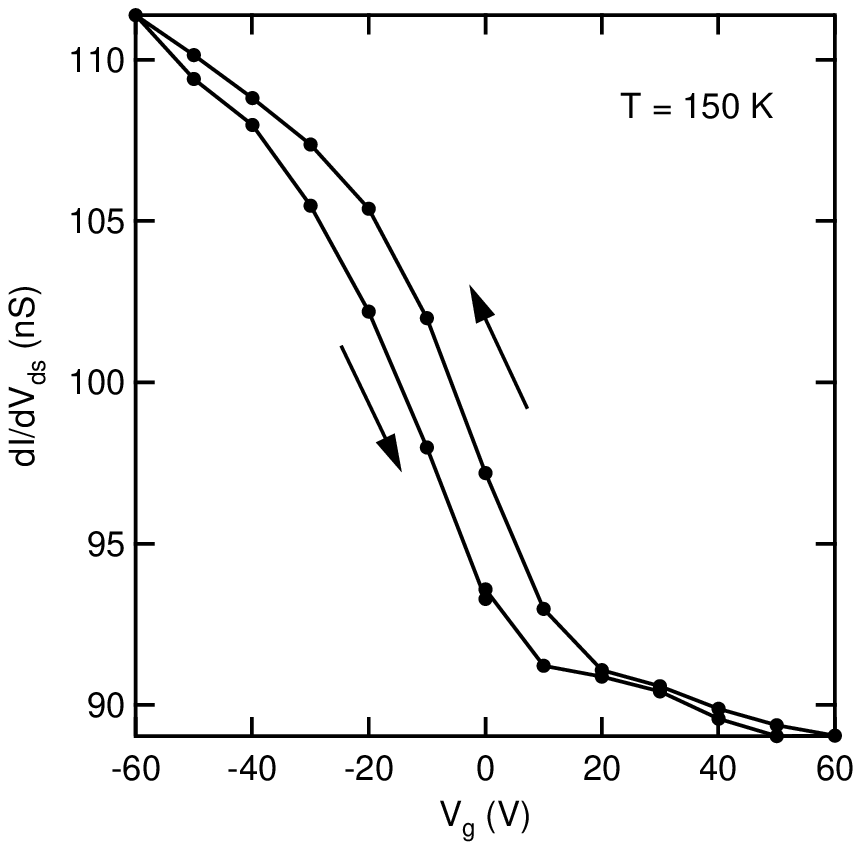}
\end{center}

\caption{Differential conductance as a function of gate voltage at \Vds\space= 0 V.  \Vg\space  is varied in increments of 10 V from 0 V to +60 V, from +60 V to -60 V, and from -60 V back to 0 V.  For each value of \Vg, the voltage is applied for 5 sec and then set to zero for a recovery period of 5 sec.  The lines serve as guides to the eye.}
\label{fig:DiffCond}
\end{figure}

We find that the current is hysteretic with variation of \Vg.  To reduce the hysteresis, we employ the following technique, in which we pulse the voltage on the gate: an AC voltage of 100 mV rms and 13 Hz is applied across the source-drain electrodes. \Vg\space is stepped in increments of 10 V.  For each value of \Vg, the potential on the gate is applied for 5 s and the differential conductance is measured. Then \Vg\space is set to 0 V for 5 s to allow the sample to relax.  This method reduces the hystersis because trapping of charge on a time scale greater than 5 s is eliminated.  Figure \ref{fig:DiffCond} displays the differential conductance as a function of \Vg\space measured in this way at 150 K.  \Vg\space is stepped from 0 V up to 60 V, then down to -60 V and back to 0 V.  These data are consistent with those shown in Fig. \ref{fig:VgDC3} in that the conductance increases (decreases) as \Vg\space is made more negative (positive).  

\begin{figure} 
\setlength{\unitlength}{1cm}
\begin{center}
\includegraphics[width=8.4cm, keepaspectratio=true]{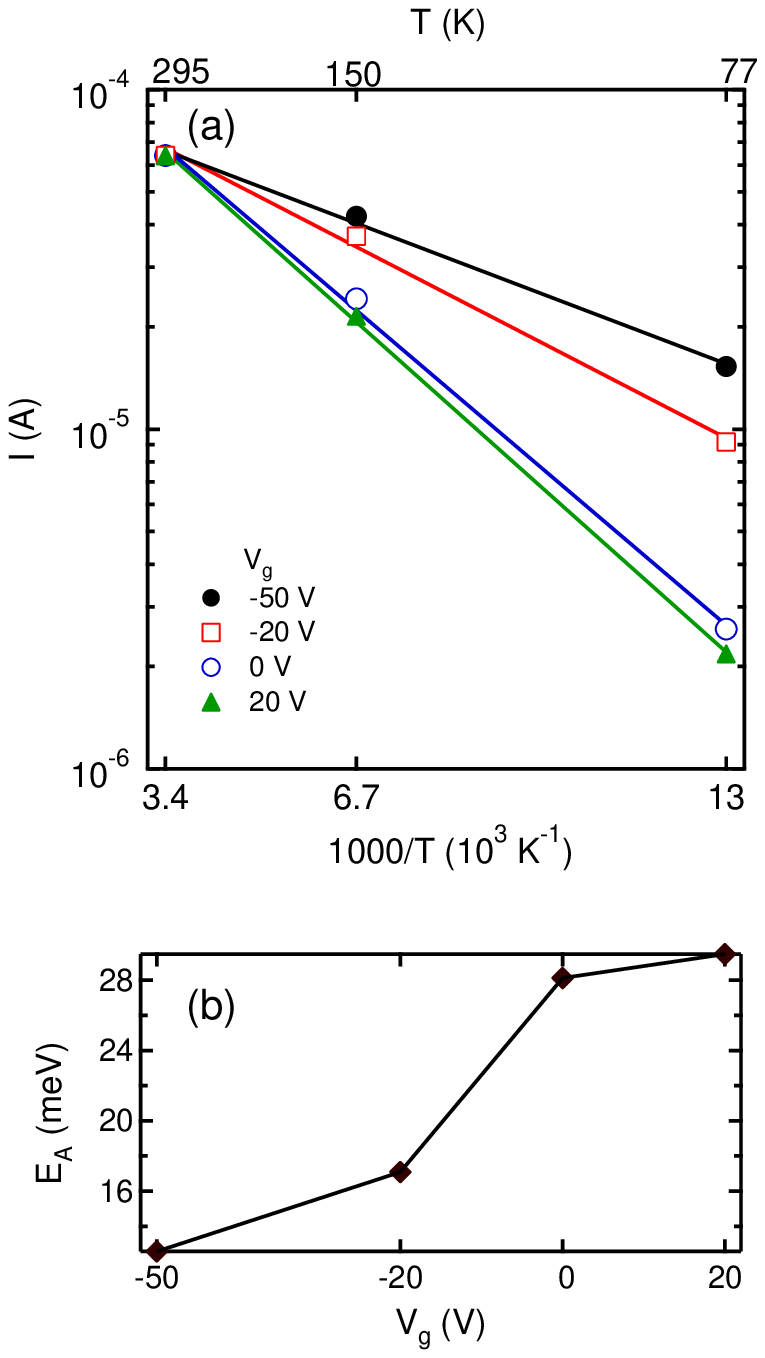}
\end{center}

\caption{(a) Temperature dependence of the current at different values of \Vg\space with \Vds\space= 6 V. The solid lines are fits to an Arrhenius equation.  (b) The activation energy, E$_{A}$, versus \Vg.  The solid line is a guide to the eye.}
\label{fig:VgDependence}
\end{figure}

The temperature dependence of the current at \Vds\space= 6 V for different values of gate voltage is displayed in Fig. \ref{fig:VgDependence}(a).  The current is measured at a fixed temperature while the gate voltage is varied and then the current is re-plotted against inverse temperature.  At each value of \Vg, the data are fit to an Arrhenius equation.  As \Vg\space is made more negative (positive), the activation energy decreases (increases), as displayed in Fig. \ref{fig:VgDependence}(b).

\section{Discussion} \label{sec:disc}

\begin{figure} 
\setlength{\unitlength}{1cm}
\begin{center}
\includegraphics[width=8.4cm, keepaspectratio=true]{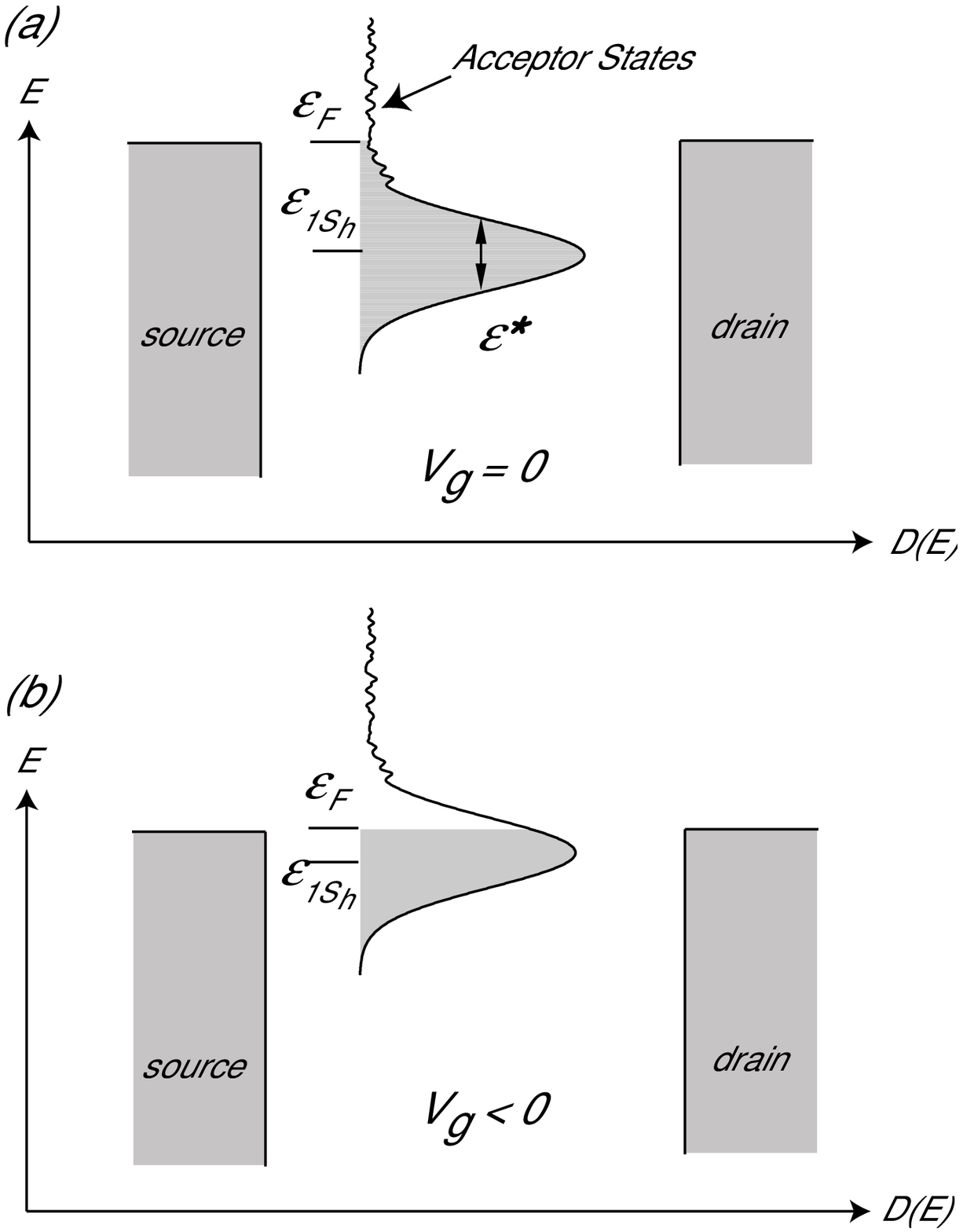}
\end{center}

\caption{Sketch of the density of states versus energy.  In this model, impurity states in the bandgap are broadly distributed in energy. \{$\varepsilon_{F} - \varepsilon_{1S_{h}}$\} is the energy required to generate a hole in the 1S$_{h}$ state. $\varepsilon$* is the width of the disorder-broadened band of 1S$_{h}$ states. (a) \Vg\space = 0 and (b) \Vg\space $<$ 0}
\label{fig:DOS}
\end{figure}

The differential conductance increases (decreases) with increasing negative (positive) \Vg\space [Fig. \ref{fig:DiffCond}], indicating that holes are the majority carrier in the NC solid as found previously.\cite{Talapin:Science}  In bulk PbSe, unpassivated Se atoms as well as oxidation create acceptor states.\cite{Bube}  It is likely that the same kinds of acceptors are present in NCs, especially at the surfaces, and that these acceptor states are distributed in energy in the bandgap.
The acceptors shift the Fermi energy toward the valence band of the NC solid. We believe, for reasons discussed below, that the density of states is similar to that sketched in Fig. \ref{fig:DOS}(a) with the the Fermi energy lying in the acceptor states that are close to the valence band.  
 
\begin{figure} 
\setlength{\unitlength}{1cm}
\begin{center}
\includegraphics[width=8.4cm, keepaspectratio=true]{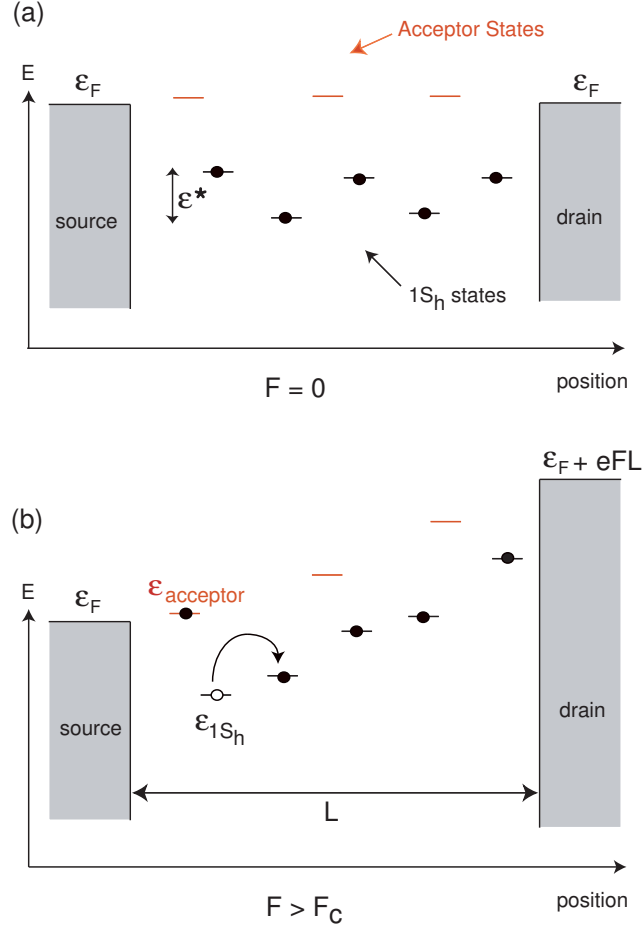}
\end{center}

\caption{(a) 1S$_{h}$ states are distributed in energy versus position with the variation given by $\varepsilon$*.  At F=0, the acceptor states right above the Fermi energy are vacant and the energy to excite a hole into the 1S$_{h}$ state is \{$\varepsilon_{F} - \varepsilon_{1S_{h}}$\}.  The subsequent energy required to hop through the array is given by $\varepsilon$*.  (b) When F $\neq$ 0, the energy of the drain electrode is raised by eFL where L is the spacing between the source and drain electrodes.  The energy to excite a hole is reduced by eFx (where x is the distance between the core and the surface of the NC.) For F $>$ F$_{c}$ (where F$_{c}$ is described in the text), holes always hop to NCs with higher site energies and thus no longer require the absorption of a phonon.  The current is limited by the energy required to excite a hole which is the difference between the energy of the acceptor state on the surface and the energy of the 1S$_{h}$ state in the NC; the latter is reduced by eFx.}
\label{fig:FieldEffect}
\end{figure}

The temperature dependence of the zero-bias conductance [Fig. \ref{fig:TempDep}] is well described by an Arrhenius equation with an activation energy E$_{A}$ = 42 meV. We propose a model in which thermal energy is required to generate a hole in the NC from the acceptor on its surface as well as for phonon-assisted hopping between NCs with different site energies.  The latter is necessary as long as the disorder in site energy is larger than the hopping matrix element between NCs. The zero bias conductivity does not distinguish between the different components of the activation energy.  Rather, the activation energy is the energy between the Fermi energy $\varepsilon_{F}$ and the highest-energy point in the percolation path that holes must traverse to pass from one electrode to the other.  For reasons that will become apparent in the discussion of the field dependence of the conductance, it is convenient to divide the activation energy as follows: \{$\varepsilon_{F}-\varepsilon_{1S_{h}}$\} is the energy required to excite a hole from an acceptor state at the Fermi energy to a 1S$_{h}$ state given at an energy of $\varepsilon_{1S_{h}}$;\footnote{Exciting a hole from the Fermi energy to the 1S$_{h}$ band is equivalent to exciting an electron from the 1S$_{h}$ band to the Fermi energy.} and, $\varepsilon$* is the energy required to hop between NCs, which is set by the variation in energy of the 1S$_{h}$ states. This is illustrated in Fig. \ref{fig:FieldEffect}(a).  Without loss of generality, we write:
\begin{equation}
E_{A} = \{\varepsilon_{F}-\varepsilon_{1S_{h}}\} + \varepsilon^{*}
\label{eq:actEngy}
\end{equation}
From optical absorption of NCs in solution, we find the variation in site energy $\varepsilon$* to be 53 meV.  [Fig. \ref{fig:AbsSAXS}(a)]. This value is consistent with previous data of P. Liljeroth \textit{et al.} who typically find that the width of the band of 1S$_{h}$ states is $\approx$65 meV in an annealed film of NCs.  While there are several possible contributions to the variation in site energy of NCs in a film, the similarity between $\varepsilon$* for NCs in a solution and in a film suggests that the dominant contribution is the variation in size of the NC and its consequent variation in confinement in energy. The activation energy at zero bias of 42 meV is smaller than these values, suggesting that the Fermi level lies in the tail of the 1S$_{h}$ band as sketched in Fig. \ref{fig:DOS}(a).    

As shown in Fig. \ref{fig:VgDC3}, the dependence of the current on source-drain voltage is essentially independent of gate voltage. This observation allows us to analyze the influence of the gate and of the source-drain voltage separately.  For a given change in gate voltage, the change in the occupancy of states and the associated change in the charge density owing to the gate is in steady state. The application of \Vds\space results in only a small variation in the steady-state charge density because \Vds\space is small compared to \Vg\space in our measurements.  Following we analyze the effect of \Vds\space on the activation energy and then the effect of \Vg.

As noted in connection with Fig. \ref{fig:VdsDependence}, the temperature dependence of the conductance changes with applied field. We expect the field to modify the energy required to generate a carrier: \{$\varepsilon_{F}-\varepsilon_{1S_{h}}$\}$\rightarrow$\{$\varepsilon_{F}-\varepsilon_{1S_{h}}$\}$-eFx$ where \textit{e} is the electron charge, \textit{F} is the electric field and \textit{x} is the distance separating an acceptor state on the surface of the NC from the 1S$_{h}$ state in the core of the NC.\footnote{Because of the random orientation of the NCs and the acceptors relative to the field, ionization will be faciliated by the field for only some of the acceptors.  However, these are the ones which will dominate the transport at high field.} The field also reduces the energy required to hop between NCs with different site energies: $\varepsilon^{*} \rightarrow \varepsilon^{*} - eFd$ where \textit{d} is the center-to-center distance between NCs.  Thus, in the presence of a field, the activation energy is given by 
\begin{equation}
E_{A} = \{\varepsilon_{F}-\varepsilon_{1S_{h}}\} - eFx + \varepsilon^{*} - eFd 
\label{eq:field}
\end{equation}
At low fields, the slope of the activation energy versus field is $e(x+d)$.  Above a critical field $F_{c}$, given by $eF_{c}d=\varepsilon$*, charge carriers always hop to NCs with lower site energies. Thus holes always hop to NCs with higher site energies as illustrated in Fig. \ref{fig:FieldEffect}(b).  Hopping no longer requires the absorption of a phonon and the slope of the activation energy versus field is simply $ex$.  These two length scales are apparent in Fig. \ref{fig:VdsDependence}(b).  At the highest fields (3$\times$10$^{4}$ V/cm $<$ F $<$ 10$^{5}$ V/cm), we find \textit{x} = 1.1$\pm$0.1 nm.  This is a measure of the distance between an acceptor state on the surface of the NC and the 1S$_{h}$ state in the core, and is a reasonable estimate of the distance between the surface of the NC and its core. In the low field region (10$^{2}$ V/cm $<$ F $<$ 10$^{4}$ V/cm), we find \textit{x+d} = 12$\pm$4 nm. Using the value \textit{x} = 1.1 nm, we find \textit{d}= 11$\pm$4 nm, which is within errors of $\approx$ 7.2 nm, the value we expect based on the size of the NCs and the thickness of the organic capping layer.  The change in the slope at $\sim\space10^{4}$ V/cm indicates that this is approximately the critical field.  At this field the activation energy has been reduced by $\approx$ 20 meV, about half the variation in energy measured optically, which seems reasonable.  At a sufficiently high field, tunneling between nanocrystals is expected to be limited by the height and width of the potential barrier caused by the molecules of the cap
layer.  However, at the highest fields used in our measurements, the current appears (Fig. \ref{fig:VdsDependence}(b)) to
still be limited by thermal release of holes from the acceptor states.

The I-\Vds\space characteristic in Fig. \ref{fig:TempFull} is consistent with this model. The conductance is ohmic at low fields (for T$\geq$40 K) and becomes non-ohmic at higher fields.  The crossover from ohmic to non-ohmic behavior is expected to occur when the field reduces the activation energy by an amount $\approx$\kboltz T.  The triangles in Fig. \ref{fig:TempFull} indicate the value of \Vds\space for which the crossover is predicted, namely where $\Delta E_{A}\approx eF(x+d) \approx\kboltz T$; and, is in good agreement with the data.

Figure \ref{fig:VgDependence} shows how the temperature dependence of the conductance changes with gate voltage at high \Vds.  As shown above, at high fields the activation energy is given by the energy required to release holes from acceptors, rather than by the energy necessary to hop between NCs.  Thus the variation in the activation energy in this regime results from a shift in the Fermi energy caused by the addition of charge to the NC solid.  Figure \ref{fig:DOS}(a) shows the position of the Fermi energy before \Vg\space is applied.  With the application of a negative \Vg, which removes electrons or adds holes, the Fermi energy shifts deeper into the band as illustrated in \ref{fig:DOS}(b). As the voltage on the gate is made more negative (positive), $\varepsilon_{F}$ decreases (increases) and the activation energy decreases (increases) as shown in Fig. \ref{fig:VgDependence}(b). The charge density added to the film by the gate is
\begin{equation}
\Delta \rho=C \Delta\Vg/s = eD(\varepsilon_{F})\Delta \varepsilon_{F}
\label{eq:addCharge}
\end{equation}
where \textit{C} is the capacitance per unit area between the gate and the film, \textit{s} is the screening length in the NC solid, \textit{e} is the electron charge and $D(\varepsilon_{F})$ is the density of states at the Fermi energy. The screening length can be estimated using the Thomas-Fermi expression
 
\begin{equation}
s = \sqrt{\frac{\kappa}{e^{2} D(\varepsilon_{F})}}
\label{eq:screening}
\end{equation}
where $\kappa$ is the dielectric constant of the NC solid.  To solve for \textit{s} from these two equations, we take $\kappa$= 250 and use the slope of the line between successive data points in Fig. \ref{fig:VgDependence}(b) for $\Delta \varepsilon_{F}/\Delta\Vg$.  We find $s = 8$ nm, approximately the size of one NC. We conclude that all the charge is added to the first monolayer of NCs adjacent to the oxide.  Previous work shows that $\kappa\geq100$ for PbSe NCs,\cite{Dielectric} but a smaller value of $\kappa$ would only make \textit{s} even smaller.  This indicates that the change in conductance as a function of gate voltage [Fig. \ref{fig:DiffCond}] results from a change in the charge density in the first monolayer only, while the conductance in the rest of the sample remains unchanged.

Knowing this, we can extract the density of states at the Fermi energy. For -20 V $<$ V$_{g}$ $<$ 0 V [Fig. \ref{fig:VgDependence}(b)], we find $D(\varepsilon_{F})\approx 10^{20}/$ eV cm$^{3}$ at \Vds\space= 6 V.  An estimate of the density of the 1S$_{h}$ states based on the linewidth of the optical absorption peak, the spatial density of the NCs and the 8-fold degeneracy of the 1S$_{h}$ state\cite{Nimtz, Kang, Murray:SynthesisII, Schaller} is $D(\varepsilon_{F})\approx4.4 \times 10^{20}$/ eV cm$^{3}$, in good agreement with the density of states derived from the data in Fig. \ref{fig:VgDependence}(b). As V$_{g}$ is made more negative than -20 V, the Fermi energy moves further into the 1S$_{h}$ band and the density of states at the Fermi energy increases as reflected in Fig. \ref{fig:DOS}(b).  Owing to the higher density of states at the Fermi energy, more holes need to be added to the film to move the Fermi energy by a given amount.  Thus a given decrease in gate voltage results in a smaller change in the Fermi energy as reflected in the decreasing slope of E$_{A}$ versus V$_{g}$ in Fig. \ref{fig:VgDependence}(b).

From our field effect data [Fig. \ref{fig:DiffCond}], we extract a field-effect hole mobility in the regime where current varies linearly with \Vds.  We find $\mu_{lin}\sim$ 4 $\times$ 10$^{-5}$ cm$^{2}$/ V s at 150 K in the regime of zero bias.  Previous studies report a hole mobility of $\mu\sim 10^{-4}$ cm$^{2}$/ V s from photocurrent measurements in arrays of CdTe NCs \cite{Porter:CdTe} and a field-effect hole mobility in chemically-treated PbSe NCs of $\mu_{lin}\sim$ 0.09 cm$^{2}$/ V s at 120 K.\cite{Talapin:Science} We find a smaller mobility most likely because our NC solids have a thicker capping layer.

Of course, our mobility is also reduced because holes are added to the NC solid at the energy of the acceptor states, not at the transport energy.  However, we estimate that this reduces the mobility by no more than a factor of $\sim$10 at 150 K [Fig. \ref{fig:DiffCond}]. One might suggest that the field-effect mobility is reduced by trapping in the oxide or at the oxide/NC-solid interface.  The hysteresis that we observe in field-effect measurments indicates that there is a large enough density of slow traps to reduce the charge added to the NC solid.  While the pulsed-gate technique eliminates trapping that occurs on a long time scale, it would not be surprising if there were also a large density of fast traps. However, the good agreement we find above between the experimentally derived and the expected density of states suggests that most of the charge is added to the NC solid. 

The model we propose ignores Coulomb interactions between holes.  On the one hand, the amount of charge added to the NC solid corresponds to of order one hole per NC in the first monolayer, and correlations are expected to be significant at these charge densities.  On the other hand, the large dielectric constant of PbSe NC solids reduces the Coulomb interaction, making it likely that disorder dominates.  Were many-body interactions strongly affecting the transport, we would expect the dependence of the current on field to change with gate voltage.  Figure \ref{fig:VgDC3} shows that there is little such change.

An alternative explanation for the small activation energy seen in the zero bias conductance is that it arises from variable-range hopping (VRH) as has been speculated.\cite{Romero:PRL, Talapin:Science, Wehrenberg}  The temperature range of Fig. \ref{fig:TempDep} is not large enough to distinguish unambiguously simply activated conduction from VRH without a Coulomb gap.  However, analysis of the data shows that if VRH is the dominant mechanism, then the hopping distance in the temperature range of our experiments is comparable to the distance between NCs. This would make the model no different from that of nearest-neighbor hopping, which we assume above.  It is also difficult to understand how others find VRH with a Coulomb gap above 4 K.\cite{Romero:PRL, Wehrenberg}  The high dielectric constant found in PbSe NCs\cite{Dielectric} indicates that a Coulomb gap would become apparent only well below 4 K.

\section{Conclusions} \label{sec:concl}

By measuring the dependence of current on source-drain voltage, gate voltage and temperature, we have shown that our PbSe NC solids are well described by a simple model of hopping between the highest energy filled states of the NCs.  After annealing, the Fermi energy is close to the edge of the highest-occupied valence level, as evinced by the small activation energy of the current and by field-effect measurements, which give a density of states close to that expected for the 1S$_{h}$ band.  

The activation energy is given by the energy to excite holes from acceptor states and activation over additional barriers resulting from the site disorder.  By applying high electric fields, we separate these two components and measure the length scales associated with them.  Not surprisingly, we find that the separation of a hole from an acceptor state occurs over a distance comparable to the spacing between the surface of the NC and its core, and the length scale associated with motion through the NC array is the center-to-center distance between NCs.  Furthermore, the energy dispersion that limits the hole motion once it is free is about the same size as the energy dispersion measured by optical spectroscopy of NCs in a colloidal suspension. The density of states extracted from field-effect data is consistent with this as well.  All of this suggests that the variation in energies of the localized intrinsic states is largely determined by the small variation in size of the NCs.

\section*{Acknowlegement} 
We are grateful to S. Amasha, I. Gelfand and I. Radu for experimental help and to S. Speakman for assistance with GISAXS. This work was funded in part by the NSF MRSEC program (DMR 0213282) at MIT and the authors made use of its shared user facilities. It was also supported by the NSEC Program of the National Science Foundation Award No. DMR-0117795 and the U.S. Army Research Office through the Institute for Soldier Nanotechnologies, under Contract No. DAAD-19-02-0002. T.S.M. gratefully acknowledges support from NDSEG.


\begin{thebibliography}{35}
\expandafter\ifx\csname natexlab\endcsname\relax\def\natexlab#1{#1}\fi
\expandafter\ifx\csname bibnamefont\endcsname\relax
  \def\bibnamefont#1{#1}\fi
\expandafter\ifx\csname bibfnamefont\endcsname\relax
  \def\bibfnamefont#1{#1}\fi
\expandafter\ifx\csname citenamefont\endcsname\relax
  \def\citenamefont#1{#1}\fi
\expandafter\ifx\csname url\endcsname\relax
  \def\url#1{\texttt{#1}}\fi
\expandafter\ifx\csname urlprefix\endcsname\relax\def\urlprefix{URL }\fi
\providecommand{\bibinfo}[2]{#2}
\providecommand{\eprint}[2][]{\url{#2}}

\bibitem[{\citenamefont{Murray et~al.}(1995)\citenamefont{Murray, Kagan, and
  Bawendi}}]{Bawendi:packing}
\bibinfo{author}{\bibfnamefont{C.~B.} \bibnamefont{Murray}},
  \bibinfo{author}{\bibfnamefont{C.~R.} \bibnamefont{Kagan}}, \bibnamefont{and}
  \bibinfo{author}{\bibfnamefont{M.~G.} \bibnamefont{Bawendi}},
  \bibinfo{journal}{Science} \textbf{\bibinfo{volume}{270}},
  \bibinfo{pages}{1335} (\bibinfo{year}{1995}).

\bibitem[{\citenamefont{Collier et~al.}(1998)\citenamefont{Collier, Vossmeyer,
  and Heath}}]{Heath:packing}
\bibinfo{author}{\bibfnamefont{C.~P.} \bibnamefont{Collier}},
  \bibinfo{author}{\bibfnamefont{T.}~\bibnamefont{Vossmeyer}},
  \bibnamefont{and} \bibinfo{author}{\bibfnamefont{J.~R.} \bibnamefont{Heath}},
  \bibinfo{journal}{Annu. Rev. Phys. Chem.} \textbf{\bibinfo{volume}{49}},
  \bibinfo{pages}{371} (\bibinfo{year}{1998}).

\bibitem[{\citenamefont{Markovitch et~al.}(1999)\citenamefont{Markovitch,
  Collier, Henrichs, Remacle, Levine, and Heath}}]{Markovitch}
\bibinfo{author}{\bibfnamefont{G.}~\bibnamefont{Markovitch}},
  \bibinfo{author}{\bibfnamefont{C.~P.} \bibnamefont{Collier}},
  \bibinfo{author}{\bibfnamefont{S.~E.} \bibnamefont{Henrichs}},
  \bibinfo{author}{\bibfnamefont{F.}~\bibnamefont{Remacle}},
  \bibinfo{author}{\bibfnamefont{R.~D.} \bibnamefont{Levine}},
  \bibnamefont{and} \bibinfo{author}{\bibfnamefont{J.~R.} \bibnamefont{Heath}},
  \bibinfo{journal}{Acc. Chem. Res.} \textbf{\bibinfo{volume}{32}},
  \bibinfo{pages}{415} (\bibinfo{year}{1999}).

\bibitem[{\citenamefont{Murray et~al.}(2001)\citenamefont{Murray, Sun,
  Gaschler, Doyle, Betley, and Kagan}}]{Murray:SynthesisII}
\bibinfo{author}{\bibfnamefont{C.~B.} \bibnamefont{Murray}},
  \bibinfo{author}{\bibfnamefont{S.}~\bibnamefont{Sun}},
  \bibinfo{author}{\bibfnamefont{W.}~\bibnamefont{Gaschler}},
  \bibinfo{author}{\bibfnamefont{H.}~\bibnamefont{Doyle}},
  \bibinfo{author}{\bibfnamefont{T.~A.} \bibnamefont{Betley}},
  \bibnamefont{and} \bibinfo{author}{\bibfnamefont{C.~R.} \bibnamefont{Kagan}},
  \bibinfo{journal}{IBM. J. Res. Dev.} \textbf{\bibinfo{volume}{45}},
  \bibinfo{pages}{47} (\bibinfo{year}{2001}).

\bibitem[{\citenamefont{Redl et~al.}(2003)\citenamefont{Redl, Cho, Murray, and
  O'Brien}}]{Redl}
\bibinfo{author}{\bibfnamefont{F.~X.} \bibnamefont{Redl}},
  \bibinfo{author}{\bibfnamefont{K.-S.} \bibnamefont{Cho}},
  \bibinfo{author}{\bibfnamefont{C.~B.} \bibnamefont{Murray}},
  \bibnamefont{and} \bibinfo{author}{\bibfnamefont{S.}~\bibnamefont{O'Brien}},
  \bibinfo{journal}{Nature} \textbf{\bibinfo{volume}{423}},
  \bibinfo{pages}{968} (\bibinfo{year}{2003}).

\bibitem[{\citenamefont{Shevchenko et~al.}(2006)\citenamefont{Shevchenko,
  Talapin, Kotov, O'Brien, and Murray}}]{Shevchenko}
\bibinfo{author}{\bibfnamefont{E.~V.} \bibnamefont{Shevchenko}},
  \bibinfo{author}{\bibfnamefont{D.~V.} \bibnamefont{Talapin}},
  \bibinfo{author}{\bibfnamefont{N.~A.} \bibnamefont{Kotov}},
  \bibinfo{author}{\bibfnamefont{S.}~\bibnamefont{O'Brien}}, \bibnamefont{and}
  \bibinfo{author}{\bibfnamefont{C.~B.} \bibnamefont{Murray}},
  \bibinfo{journal}{Nature} \textbf{\bibinfo{volume}{439}}, \bibinfo{pages}{55}
  (\bibinfo{year}{2006}).

\bibitem[{\citenamefont{Whan et~al.}(1996)\citenamefont{Whan, White, and
  Orlando}}]{Orlando}
\bibinfo{author}{\bibfnamefont{C.~B.} \bibnamefont{Whan}},
  \bibinfo{author}{\bibfnamefont{J.}~\bibnamefont{White}}, \bibnamefont{and}
  \bibinfo{author}{\bibfnamefont{T.~P.} \bibnamefont{Orlando}},
  \bibinfo{journal}{Appl. Phys. Lett.} \textbf{\bibinfo{volume}{68}},
  \bibinfo{pages}{2996} (\bibinfo{year}{1996}).

\bibitem[{\citenamefont{Talapin and Murray}(2005)}]{Talapin:Science}
\bibinfo{author}{\bibfnamefont{D.~V.} \bibnamefont{Talapin}} \bibnamefont{and}
  \bibinfo{author}{\bibfnamefont{C.~B.} \bibnamefont{Murray}},
  \bibinfo{journal}{Science} \textbf{\bibinfo{volume}{310}},
  \bibinfo{pages}{86} (\bibinfo{year}{2005}).

\bibitem[{\citenamefont{Novikov et~al.}(2005)\citenamefont{Novikov, Kozinsky,
  and Levitov}}]{Levitov:theory}
\bibinfo{author}{\bibfnamefont{D.~S.} \bibnamefont{Novikov}},
  \bibinfo{author}{\bibfnamefont{B.}~\bibnamefont{Kozinsky}}, \bibnamefont{and}
  \bibinfo{author}{\bibfnamefont{L.~S.} \bibnamefont{Levitov}},
  \bibinfo{journal}{Phys. Rev. B} \textbf{\bibinfo{volume}{72}},
  \bibinfo{pages}{235331} (\bibinfo{year}{2005}).

\bibitem[{\citenamefont{Morgan et~al.}(2002)\citenamefont{Morgan, Leatherdale,
  {Drndi\'{c}}, Jarosz, Kastner, and Bawendi}}]{Morgan:CdSe}
\bibinfo{author}{\bibfnamefont{N.~Y.} \bibnamefont{Morgan}},
  \bibinfo{author}{\bibfnamefont{C.~A.} \bibnamefont{Leatherdale}},
  \bibinfo{author}{\bibfnamefont{M.}~\bibnamefont{{Drndi\'{c}}}},
  \bibinfo{author}{\bibfnamefont{M.~V.} \bibnamefont{Jarosz}},
  \bibinfo{author}{\bibfnamefont{M.~A.} \bibnamefont{Kastner}},
  \bibnamefont{and} \bibinfo{author}{\bibfnamefont{M.~G.}
  \bibnamefont{Bawendi}}, \bibinfo{journal}{Phys. Rev. B}
  \textbf{\bibinfo{volume}{66}}, \bibinfo{pages}{075339}
  (\bibinfo{year}{2002}).

\bibitem[{\citenamefont{{Drndi\'{c}} et~al.}(2002)\citenamefont{{Drndi\'{c}},
  Jarosz, Morgan, Kastner, and Bawendi}}]{Drndic:Anneal}
\bibinfo{author}{\bibfnamefont{M.}~\bibnamefont{{Drndi\'{c}}}},
  \bibinfo{author}{\bibfnamefont{M.~V.} \bibnamefont{Jarosz}},
  \bibinfo{author}{\bibfnamefont{N.~Y.} \bibnamefont{Morgan}},
  \bibinfo{author}{\bibfnamefont{M.~A.} \bibnamefont{Kastner}},
  \bibnamefont{and} \bibinfo{author}{\bibfnamefont{M.~G.}
  \bibnamefont{Bawendi}}, \bibinfo{journal}{J. Appl. Phys.}
  \textbf{\bibinfo{volume}{92}}, \bibinfo{pages}{7498} (\bibinfo{year}{2002}).

\bibitem[{\citenamefont{Ginger and Greenham}(2000)}]{Ginger}
\bibinfo{author}{\bibfnamefont{D.~S.} \bibnamefont{Ginger}} \bibnamefont{and}
  \bibinfo{author}{\bibfnamefont{N.~C.} \bibnamefont{Greenham}},
  \bibinfo{journal}{J. Appl. Phys.} \textbf{\bibinfo{volume}{87}},
  \bibinfo{pages}{1361} (\bibinfo{year}{2000}).

\bibitem[{\citenamefont{Fischbein and Drndi\'{c}}(2005)}]{Fischbein}
\bibinfo{author}{\bibfnamefont{M.~D.} \bibnamefont{Fischbein}}
  \bibnamefont{and}
  \bibinfo{author}{\bibfnamefont{M.}~\bibnamefont{Drndi\'{c}}},
  \bibinfo{journal}{Appl. Phys. Lett.} \textbf{\bibinfo{volume}{86}},
  \bibinfo{pages}{193106} (\bibinfo{year}{2005}).

\bibitem[{\citenamefont{Drndi\'{c} et~al.}(2003)\citenamefont{Drndi\'{c},
  Markov, Jarosz, Bawendi, and Kastner}}]{DrndicII}
\bibinfo{author}{\bibfnamefont{M.}~\bibnamefont{Drndi\'{c}}},
  \bibinfo{author}{\bibfnamefont{R.}~\bibnamefont{Markov}},
  \bibinfo{author}{\bibfnamefont{M.~V.} \bibnamefont{Jarosz}},
  \bibinfo{author}{\bibfnamefont{M.~G.} \bibnamefont{Bawendi}},
  \bibnamefont{and} \bibinfo{author}{\bibfnamefont{M.~A.}
  \bibnamefont{Kastner}}, \bibinfo{journal}{Appl. Phys. Lett.}
  \textbf{\bibinfo{volume}{83}}, \bibinfo{pages}{4008} (\bibinfo{year}{2003}).

\bibitem[{\citenamefont{Leatherdale et~al.}(2000)\citenamefont{Leatherdale,
  Kagan, Morgan, Empedocles, Kastner, and Bawendi}}]{Leatherdale}
\bibinfo{author}{\bibfnamefont{C.~A.} \bibnamefont{Leatherdale}},
  \bibinfo{author}{\bibfnamefont{C.~R.} \bibnamefont{Kagan}},
  \bibinfo{author}{\bibfnamefont{N.~Y.} \bibnamefont{Morgan}},
  \bibinfo{author}{\bibfnamefont{S.~A.} \bibnamefont{Empedocles}},
  \bibinfo{author}{\bibfnamefont{M.~A.} \bibnamefont{Kastner}},
  \bibnamefont{and} \bibinfo{author}{\bibfnamefont{M.~G.}
  \bibnamefont{Bawendi}}, \bibinfo{journal}{Phys. Rev. B}
  \textbf{\bibinfo{volume}{62}}, \bibinfo{pages}{2669} (\bibinfo{year}{2000}).

\bibitem[{\citenamefont{Allan and Delerue}(2004)}]{AllanDelerue}
\bibinfo{author}{\bibfnamefont{G.}~\bibnamefont{Allan}} \bibnamefont{and}
  \bibinfo{author}{\bibfnamefont{C.}~\bibnamefont{Delerue}},
  \bibinfo{journal}{Phys. Rev. B} \textbf{\bibinfo{volume}{70}},
  \bibinfo{pages}{245321} (\bibinfo{year}{2004}).

\bibitem[{\citenamefont{Wehrenberg et~al.}(2002)\citenamefont{Wehrenberg, Wang,
  and Guyot-Sionnest}}]{Optical}
\bibinfo{author}{\bibfnamefont{B.~L.} \bibnamefont{Wehrenberg}},
  \bibinfo{author}{\bibfnamefont{C.}~\bibnamefont{Wang}}, \bibnamefont{and}
  \bibinfo{author}{\bibfnamefont{P.}~\bibnamefont{Guyot-Sionnest}},
  \bibinfo{journal}{J. Phys. Chem. B.} \textbf{\bibinfo{volume}{106}},
  \bibinfo{pages}{10634} (\bibinfo{year}{2002}).

\bibitem[{\citenamefont{Schaller and Klimov}(2004)}]{Klimov}
\bibinfo{author}{\bibfnamefont{R.~D.} \bibnamefont{Schaller}} \bibnamefont{and}
  \bibinfo{author}{\bibfnamefont{V.~I.} \bibnamefont{Klimov}},
  \bibinfo{journal}{Phys. Rev. Lett.} \textbf{\bibinfo{volume}{92}},
  \bibinfo{pages}{186601} (\bibinfo{year}{2004}).

\bibitem[{\citenamefont{Coe et~al.}(2002)\citenamefont{Coe, Woo, Bawendi, and
  Bulovic}}]{LED}
\bibinfo{author}{\bibfnamefont{S.}~\bibnamefont{Coe}},
  \bibinfo{author}{\bibfnamefont{W.-K.} \bibnamefont{Woo}},
  \bibinfo{author}{\bibfnamefont{M.~G.} \bibnamefont{Bawendi}},
  \bibnamefont{and} \bibinfo{author}{\bibfnamefont{V.}~\bibnamefont{Bulovic}},
  \bibinfo{journal}{Nature} \textbf{\bibinfo{volume}{420}},
  \bibinfo{pages}{800} (\bibinfo{year}{2002}).

\bibitem[{\citenamefont{Huynh et~al.}(2002)\citenamefont{Huynh, Dittmer, and
  Alivisatos}}]{AlivisatosSolar}
\bibinfo{author}{\bibfnamefont{W.~U.} \bibnamefont{Huynh}},
  \bibinfo{author}{\bibfnamefont{J.~J.} \bibnamefont{Dittmer}},
  \bibnamefont{and} \bibinfo{author}{\bibfnamefont{A.~P.}
  \bibnamefont{Alivisatos}}, \bibinfo{journal}{Science}
  \textbf{\bibinfo{volume}{295}}, \bibinfo{pages}{2425} (\bibinfo{year}{2002}).

\bibitem[{\citenamefont{Konstantatos et~al.}(2006)\citenamefont{Konstantatos,
  Howard, Fischer, Hoogland, Clifford, Klem, Levina, and Sargent}}]{Sargent}
\bibinfo{author}{\bibfnamefont{G.}~\bibnamefont{Konstantatos}},
  \bibinfo{author}{\bibfnamefont{I.}~\bibnamefont{Howard}},
  \bibinfo{author}{\bibfnamefont{A.}~\bibnamefont{Fischer}},
  \bibinfo{author}{\bibfnamefont{S.}~\bibnamefont{Hoogland}},
  \bibinfo{author}{\bibfnamefont{J.}~\bibnamefont{Clifford}},
  \bibinfo{author}{\bibfnamefont{E.}~\bibnamefont{Klem}},
  \bibinfo{author}{\bibfnamefont{L.}~\bibnamefont{Levina}}, \bibnamefont{and}
  \bibinfo{author}{\bibfnamefont{E.~H.} \bibnamefont{Sargent}},
  \bibinfo{journal}{Nature} \textbf{\bibinfo{volume}{442}},
  \bibinfo{pages}{180} (\bibinfo{year}{2006}).

\bibitem[{\citenamefont{Romero and {Drndi\'{c}}}(2005)}]{Romero:PRL}
\bibinfo{author}{\bibfnamefont{H.~E.} \bibnamefont{Romero}} \bibnamefont{and}
  \bibinfo{author}{\bibfnamefont{M.}~\bibnamefont{{Drndi\'{c}}}},
  \bibinfo{journal}{Phys. Rev. Lett.} \textbf{\bibinfo{volume}{95}},
  \bibinfo{pages}{156801} (\bibinfo{year}{2005}).

\bibitem[{\citenamefont{Wehrenberg et~al.}(2005)\citenamefont{Wehrenberg, Yu,
  Ma, and Guyot-Sionnest}}]{Wehrenberg}
\bibinfo{author}{\bibfnamefont{B.~L.} \bibnamefont{Wehrenberg}},
  \bibinfo{author}{\bibfnamefont{D.}~\bibnamefont{Yu}},
  \bibinfo{author}{\bibfnamefont{J.}~\bibnamefont{Ma}}, \bibnamefont{and}
  \bibinfo{author}{\bibfnamefont{P.}~\bibnamefont{Guyot-Sionnest}},
  \bibinfo{journal}{J. Phys. Chem. B} \textbf{\bibinfo{volume}{109}},
  \bibinfo{pages}{20192} (\bibinfo{year}{2005}).

\bibitem[{\citenamefont{Wehrenberg and
  Guyot-Sionnest}(2003)}]{Wehrenberg:ChargeInjection}
\bibinfo{author}{\bibfnamefont{B.~L.} \bibnamefont{Wehrenberg}}
  \bibnamefont{and}
  \bibinfo{author}{\bibfnamefont{P.}~\bibnamefont{Guyot-Sionnest}},
  \bibinfo{journal}{J. Am. Chem. Soc.} \textbf{\bibinfo{volume}{125}},
  \bibinfo{pages}{7806} (\bibinfo{year}{2003}).

\bibitem[{\citenamefont{Chen et~al.}(2002)\citenamefont{Chen, Stokes, Zhou,
  Fang, and Murray}}]{Murray:Synthesis}
\bibinfo{author}{\bibfnamefont{F.}~\bibnamefont{Chen}},
  \bibinfo{author}{\bibfnamefont{K.~L.} \bibnamefont{Stokes}},
  \bibinfo{author}{\bibfnamefont{W.}~\bibnamefont{Zhou}},
  \bibinfo{author}{\bibfnamefont{J.}~\bibnamefont{Fang}}, \bibnamefont{and}
  \bibinfo{author}{\bibfnamefont{C.~B.} \bibnamefont{Murray}},
  \bibinfo{journal}{Mat. Res. Soc. Symp. Proc.} \textbf{\bibinfo{volume}{691}},
  \bibinfo{pages}{G.10.2.1} (\bibinfo{year}{2002}).

\bibitem[{\citenamefont{Steckel et~al.}(2003)\citenamefont{Steckel,
  Coe-Sullivan, Bulovi\'{c}, and Bawendi}}]{Steckel}
\bibinfo{author}{\bibfnamefont{J.~S.} \bibnamefont{Steckel}},
  \bibinfo{author}{\bibfnamefont{S.}~\bibnamefont{Coe-Sullivan}},
  \bibinfo{author}{\bibfnamefont{V.}~\bibnamefont{Bulovi\'{c}}},
  \bibnamefont{and} \bibinfo{author}{\bibfnamefont{M.~G.}
  \bibnamefont{Bawendi}}, \bibinfo{journal}{Adv. Mater.}
  \textbf{\bibinfo{volume}{15}}, \bibinfo{pages}{1862} (\bibinfo{year}{2003}).

\bibitem[{\citenamefont{Porter et~al.}(2006)\citenamefont{Porter, Mentzel,
  Charpentier, Kastner, and Bawendi}}]{Porter:CdTe}
\bibinfo{author}{\bibfnamefont{V.~J.} \bibnamefont{Porter}},
  \bibinfo{author}{\bibfnamefont{T.}~\bibnamefont{Mentzel}},
  \bibinfo{author}{\bibfnamefont{S.}~\bibnamefont{Charpentier}},
  \bibinfo{author}{\bibfnamefont{M.~A.} \bibnamefont{Kastner}},
  \bibnamefont{and} \bibinfo{author}{\bibfnamefont{M.~G.}
  \bibnamefont{Bawendi}}, \bibinfo{journal}{Phys. Rev. B}
  \textbf{\bibinfo{volume}{73}}, \bibinfo{pages}{155303}
  (\bibinfo{year}{2006}).

\bibitem[{\citenamefont{Kang and Wise}(1997)}]{Kang}
\bibinfo{author}{\bibfnamefont{I.}~\bibnamefont{Kang}} \bibnamefont{and}
  \bibinfo{author}{\bibfnamefont{F.~W.} \bibnamefont{Wise}},
  \bibinfo{journal}{J. Opt. Soc. Am. B} \textbf{\bibinfo{volume}{14}},
  \bibinfo{pages}{1632} (\bibinfo{year}{1997}).

\bibitem[{\citenamefont{Du et~al.}(2002)\citenamefont{Du, Chen, Krishnan,
  Krauss, Harbold, Wise, Thomas, and Silcox}}]{Du}
\bibinfo{author}{\bibfnamefont{H.}~\bibnamefont{Du}},
  \bibinfo{author}{\bibfnamefont{C.}~\bibnamefont{Chen}},
  \bibinfo{author}{\bibfnamefont{R.}~\bibnamefont{Krishnan}},
  \bibinfo{author}{\bibfnamefont{T.}~\bibnamefont{Krauss}},
  \bibinfo{author}{\bibfnamefont{J.~M.} \bibnamefont{Harbold}},
  \bibinfo{author}{\bibfnamefont{F.~W.} \bibnamefont{Wise}},
  \bibinfo{author}{\bibfnamefont{M.~G.} \bibnamefont{Thomas}},
  \bibnamefont{and} \bibinfo{author}{\bibfnamefont{J.}~\bibnamefont{Silcox}},
  \bibinfo{journal}{Nano Letters} \textbf{\bibinfo{volume}{2}},
  \bibinfo{pages}{1321} (\bibinfo{year}{2002}).

\bibitem[{\citenamefont{Porter et~al.}(2007)\citenamefont{Porter, Geyer,
  Halpert, Mentzel, Kastner, and Bawendi}}]{Scott}
\bibinfo{author}{\bibfnamefont{V.~J.} \bibnamefont{Porter}},
  \bibinfo{author}{\bibfnamefont{S.}~\bibnamefont{Geyer}},
  \bibinfo{author}{\bibfnamefont{J.~E.} \bibnamefont{Halpert}},
  \bibinfo{author}{\bibfnamefont{T.~S.} \bibnamefont{Mentzel}},
  \bibinfo{author}{\bibfnamefont{M.~A.} \bibnamefont{Kastner}},
  \bibnamefont{and} \bibinfo{author}{\bibfnamefont{M.~G.}
  \bibnamefont{Bawendi}} (\bibinfo{year}{2007}), \bibinfo{note}{unpublished}.

\bibitem[{\citenamefont{Bube}(1960)}]{Bube}
\bibinfo{author}{\bibfnamefont{R.}~\bibnamefont{Bube}},
  \emph{\bibinfo{title}{Photoconductivity of Solids}}
  (\bibinfo{publisher}{Wiley}, \bibinfo{year}{1960}).

\bibitem[{\citenamefont{Liljeroth et~al.}(2006)\citenamefont{Liljeroth,
  Overgaag, Urbieta, Grandidier, Hickey, and Vanmaekelbergh}}]{Liljeroth:STM}
\bibinfo{author}{\bibfnamefont{P.}~\bibnamefont{Liljeroth}},
  \bibinfo{author}{\bibfnamefont{K.}~\bibnamefont{Overgaag}},
  \bibinfo{author}{\bibfnamefont{A.}~\bibnamefont{Urbieta}},
  \bibinfo{author}{\bibfnamefont{B.}~\bibnamefont{Grandidier}},
  \bibinfo{author}{\bibfnamefont{S.~G.} \bibnamefont{Hickey}},
  \bibnamefont{and}
  \bibinfo{author}{\bibfnamefont{D.}~\bibnamefont{Vanmaekelbergh}},
  \bibinfo{journal}{Phys. Rev. Lett.} \textbf{\bibinfo{volume}{97}},
  \bibinfo{eid}{096803} (\bibinfo{year}{2006}).

\bibitem[{\citenamefont{Ben-Porat et~al.}(2004)\citenamefont{Ben-Porat,
  Chernyavskaya, Brus, Cho, and Murray}}]{Dielectric}
\bibinfo{author}{\bibfnamefont{C.~H.} \bibnamefont{Ben-Porat}},
  \bibinfo{author}{\bibfnamefont{O.}~\bibnamefont{Chernyavskaya}},
  \bibinfo{author}{\bibfnamefont{L.}~\bibnamefont{Brus}},
  \bibinfo{author}{\bibfnamefont{K.-S.} \bibnamefont{Cho}}, \bibnamefont{and}
  \bibinfo{author}{\bibfnamefont{C.~B.} \bibnamefont{Murray}},
  \bibinfo{journal}{J. Phys. Chem. B} \textbf{\bibinfo{volume}{108}},
  \bibinfo{pages}{7814} (\bibinfo{year}{2004}).

\bibitem[{\citenamefont{Nimtz and Schlicht}(1983)}]{Nimtz}
\bibinfo{author}{\bibfnamefont{G.}~\bibnamefont{Nimtz}} \bibnamefont{and}
  \bibinfo{author}{\bibfnamefont{B.}~\bibnamefont{Schlicht}}, in
  \emph{\bibinfo{booktitle}{Narrow-Gap Semiconductors}}, edited by
  \bibinfo{editor}{\bibfnamefont{G.}~\bibnamefont{H\"ohler}}
  (\bibinfo{publisher}{Springer-Verlag}, \bibinfo{year}{1983}), pp.
  \bibinfo{pages}{1--117}.

\bibitem[{\citenamefont{Shcaller et~al.}(2003)\citenamefont{Shcaller, Petruska,
  and Klimov}}]{Schaller}
\bibinfo{author}{\bibfnamefont{R.}~\bibnamefont{Shcaller}},
  \bibinfo{author}{\bibfnamefont{M.~A.} \bibnamefont{Petruska}},
  \bibnamefont{and} \bibinfo{author}{\bibfnamefont{V.~I.}
  \bibnamefont{Klimov}}, \bibinfo{journal}{J. Phys. Chem. B}
  \textbf{\bibinfo{volume}{107}}, \bibinfo{pages}{13765}
  (\bibinfo{year}{2003}).

\end{thebibliography}

\end{document}